\documentstyle[amsbsy]{article}
\makeatletter

          \@addtoreset{equation}{section}
       \makeatother

\newcommand{\nonum}{\nonumber\\}
\newcommand{\be}{\begin{equation}}
\newcommand{\ee}{\end{equation}}
\newcommand{\bea}{\begin{eqnarray}}
\newcommand{\eea}{\end{eqnarray}}
\newcommand{\bd}{\begin{displaymath}}
\newcommand{\ed}{\end{displaymath}}

\newcommand{\tr}{\hbox{\rm Tr}\,}

\begin{document}
\title{%
Generalized Chern-Simons Form and Descent Equation
}
\author{%
Yoshitaka {\sc Okumura}
\thanks{On leave of absence from Chubu University, Kasugai, 487-8501, Japan}
\thanks{
e-mail address: okum@bu.edu}\\
{\it Department of Physics, 
Boston University, Boston, MA 02215
}}
\date{}
\maketitle
\vskip 0.5cm
\begin{abstract}
{We present the general method to introduce the generalized Chern-Simons form 
and the descent equation 
which contain the scalar field in addition to the gauge fields. 
It is based on the technique in a noncommutative differential geometry (NCG)
which extends the 4-dimensional Minkowski space $M_4$ 
to the discrete space such as 
$M_4\times Z_2$ with two point space $Z_2$.
However, the resultant equations do not depend on NCG but are justified by
the algebraic rules in the ordinary differential geometry. 
\vskip 0.2cm}
\end{abstract}
\thispagestyle{empty}
\section{Introduction}
The Chern-Simons theory \cite{ChernSimons1} has been extensively
 studied so far 
with great interests both for their theoretical interests 
as the topological quantum field theories \cite{Witten}
and their practical applications for 
certain planar condensed matter phenomena
such as the fractional quantum Hall effects and high temperature 
super conductivity \cite{Wilczek}, \cite{Lerda}.
Especially, three dimensional Chern-Simons theory 
depending on three dimensional Chern-Simons form \cite{ChernSimons} provides 
a field theoretic framework for studying knots and links in three
dimension.
Furthermore, three dimensional gravity with a negative cosmological
constant is described by two Chern-Simons theories \cite{Witten2}. 
This approach \cite{Strominger} makes it possible 
to exactly calculate the black hole entropy
beyond the semi-classical calculations.
\par
The occurrences of Yang-Mills anomalies and other topological terms such as 
axial anomaly, Schwinger terms and Chern characters are the important aspect
of quantized gauge theories. Thus, the descent equations \cite{Zumino} 
are very important because a series of these equations prescribe 
the relations between Yang-Mills anomalies.
\par
Connes proposed the noncommutative geometry on the product space of
the 4-dimensional Minkowski space \cite{Connes} and two point space $Z_2$. 
The Higgs boson field is regarded as a kind of gauge field 
on the discrete space $Z_2$ in this formulation.
In fact, the Higgs boson has several similarities with the ordinary gauge
fields such as the same type couplings with fermions and the trilinear
and quartic self-couplings.
The Higgs mechanism naturally works without assuming the Higgs potential
leading to the spontaneous breakdown of gauge symmetry.
\par
After the original formulation of NCG by Connes \cite{Connes}, 
many versions of NCG  \cite{MM} 
has appeared and succeeded to reconstruct the 
spontaneously broken gauge theories.
Morita and the present author \cite{MO1} proposed the generalized 
differential geometry (GDG) on the discrete space $M_4\times Z_2$
and reconstructed the Weinberg-Salam model. 
In this formulation on $M_4\times Z_2$ 
the extra differential one-form $\chi$ 
is introduced in addition to
the usual one-form $dx^\mu$ and so our formalism is the generalization
of the ordinary differential geometry on the continuous manifold.
This formulation was generalized to GDG on the discrete space  
$M_4\times Z_{N}$ \cite{GUT}, \cite{LR} 
by introducing the extra one-forms $\chi_k(k=1,2\cdots N)$,
which generalization enabled us to reconstruct the left-right 
symmetric gauge theory, SU(5) GUT and SO(10) GUT.
\par
From the standpoint of NCG, the Higgs boson is a gauge field of 
the principal bundle on the discrete space. Thus, it is expected 
that there exist the Chern-Simons forms and descent equations 
including the scalar boson field 
in addition to the ordinary gauge field. In this letter, we address 
this problem and present the general method to introduce these generalized
Chern-Simons form and descent equations 
by use of the technique in NCG. It should be noted that
we use NCG but the resultant formulas are free from NCG and are justified
by the direct calculations in the ordinary differential geometry.
\par
\section{Differential geometry on the product space 
${\boldsymbol {M_N\times Z_2}}$}
The generalized gauge field is defined on the product space 
$M_N\times Z_2$ with $N$-dimensional Minkowski space $M_N$ and the 
two points space $Z_2$ as 
\be
{\cal A}(x)=\left(
\begin{array}{cc}
  A_1(x) & H_{12}(x)\chi_2 \\
  H_{21}(x)\chi_1 & A_2(x)
\end{array}
\right),
\label{2.1}
\ee
where
$A_1(x)\left(=-A_1^\dagger(x)\right)$ 
and $A_2(x)\left(=-A_2^\dagger(x)\right)$ 
are gauge fields belonging to the self-adjoint
representations of unitary gauge groups $G_1$ and $G_2$, respectively and 
$H_{12}(x)\left(=H_{21}(x)^\dagger\right)$ is a scalar field 
belonging to the covariant representation
of $G_1$ and $G_2$. We do not call $H_{12}$ the Higgs boson field because
its vacuum expectation value is irrelevant to our formulation. 
In addition, 
$A_1(x)=A_1^\mu(x)dx_\mu$, $A_2(x)=A_2^\mu(x)dx_\mu$ and 
$\chi_1$ and $\chi_2$ are one-form base on the discrete space $Z_2$
which satisfy the following algebraic rules.
\be
dx_\mu\wedge dx_\nu=-dx_\nu\wedge dx_\mu,
\hskip0.5cm
dx_\mu\wedge \chi_k=-\chi_k\wedge dx_\mu,
\hskip0.5cm
\chi_k\wedge\chi_l=-\chi_l\wedge\chi_k,
\label{2.2}
\ee
with $k,l= 1, 2$. We abbreviate the argument $x$ in the field hereafter
except for the case necessary to write. 
Let us first address the gauge transformation of $\cal A$ with the 
gauge function 
\be
   g=\left(\matrix{g_1 & 0 \cr
                   0   & g_2\cr}
     \right),
\label{2.3}
\ee
where $g_1\in G_1$ and $g_2\in G_2$.
It is expressed as
\be
      {\cal A}^g=g^{-1}{d} g + g^{-1}{\cal A}g,
\label{2.4}
\ee
where the operator $d=\partial^\mu dx_\mu$ is the exterior derivative
on the space $M_N$.
Equation (\ref{2.3}) brings the gauge transformations of gauge and scalar 
fields.
\bea
 && A_i^g=g_i^{-1}dg_i+g_i^{-1}A_ig_i, \nonum
 && H_{12}^g=g_1^{-1}H_{12}g_2\,. 
\label{2.5}
\eea
\par
The generalized field strength $\cal F$ is defined as usual and expressed as
\be
  {\cal F}=d{\cal A}+{\cal A}\wedge{\cal A},
\label{2.6} 
\ee
which is written in components as
\be
{\cal F}=\left(
               \matrix{ F_1+H_{12}H_{21}\chi_2\wedge\chi_1 
                            & {\cal D}H_{12}\chi_2\cr
                           {\cal D}H_{21}\chi_1  &
                         F_2+H_{21}H_{12}\chi_1\wedge\chi_2}
         \right), 
\label{2.7}
\ee
where
$F_1$ and $F_2$ are the field strength of gauge fields $A_1$ and $A_2$,
respectively and ${\cal D}H_{12}$ and ${\cal D}H_{21}$ 
are the covariant derivatives of the scalar field. Equations of 
those quantities are written as
\bea
&&   F_i=dA_i+A_i\wedge A_i, \nonum
&&   {\cal D}H_{12}=dH_{12}+A_1H_{12}-H_{12}A_2, \nonum
&&   {\cal D}H_{21}=dH_{21}+A_2H_{21}-H_{21}A_1.
\label{2.8}
\eea
According to Eq.(\ref{2.4}), we can easily find the generalized
field strength transformed covariantly under the gauge transformation;
\be
   {\cal F}^g=g^{-1}{\cal F}g,
\label{2.9}
\ee
which is written in components as
\be
   F_i^g=g_i^{-1}F_ig_i,\hskip0.5cm
   {\cal D}H_{12}^g=g_1^{-1}{\cal D}H_{12}g_2,\hskip0.5cm
   {\cal D}H_{21}^g=g_2^{-1}{\cal D}H_{21}g_1.   
\label{2.10}
\ee
The generalized field strength defined in
Eq.(\ref{2.6}) satisfies the Bianchi Identity 
\be
     {\cal D}{\cal F}=d{\cal}{\cal F}+{\cal A}{\cal F}-{\cal F}{\cal A}=0,
\label{2.11}
\ee
which is easily proved by use of the algebraic rule in the differential 
geometry and very important frequently used hereafter.
It should be noted that the Bianchi Identity in Eq.(\ref{2.10})
does not yield any restriction between gauge field $A_i$ and scalar
field $H_{12}$. 
\section{Generalized Chern-Simons form}
In order to introduce the generalized Chern-Simons form, we use the 
Cartan's homotopy formula
\be
P({\cal A}, {\cal F})=(kd+dk)P({\cal A}, {\cal F}),
\label{3.1}
\ee
where
$P({\cal A}, {\cal F})$ is an arbitrary function of ${\cal A}$ and ${\cal F}$.
The operator $k$ is defined through the equation
\be
kP({\cal A}, {\cal F})=\int_0^1dt k_tP({\cal A}_t, {\cal F}_t),
\label{3.2}
\ee
where ${\cal A}_t=t{\cal A}$ and 
${\cal F}_t=td{\cal A}+t^2{\cal A}\wedge{\cal A}$. The operator
$k_t$ in Eq.(\ref{3.2}) is an anti-differential operator and 
is defined as 
\be
  k_t{\cal A}_t=0, \hskip0.5cm k_t{\cal F}_t=t{\cal A},
\label{3.3}
\ee
through which the identity
\be
    \frac{\partial}{\partial t}=k_td+dk_t
\label{3.4}
\ee
follows. Equation (\ref{3.4}) justifies the Cartan's homotopy formula
together with Eq.(\ref{3.2}).
\par
It is easily see that the quantity $\tr {\cal{F}}^n$ is invariant
under the gauge transformation and satisfies the equation
\be
 d\tr {\cal{F}}^n=\tr \left(d{\cal{F}}^n
 +{\cal A}{\cal{F}}^n-{\cal{F}}^n{\cal A}\right)=0,
\label{3.5}
\ee
which is proved from the Bianchi Identity Eq. (\ref{2.11}).
Putting $P({\cal A}, {\cal F})=\tr {\cal{F}}^{n+1}$ in Eq.(\ref{3.1}), 
we find the transgression formula \cite{ChernSimons}
\be
 \tr {\cal{F}}^{n+1}=d\omega_{2n+1}({\cal A}, {\cal F}),
\label{3.6}
\ee
where $\omega_{2n+1}({\cal A}, {\cal F})$ is the generalized Chern-Simons
form and written as
\be
\omega_{2n+1}({\cal A}, {\cal F})=k\tr {\cal F}^{n+1}=
(n+1)\int_0^1dt  {\cal A}{\cal F}_t^n.
\label{3.7}
\ee
Paying attention on ${\cal F}_t=t{\cal F}+(t^2-t){\cal A}^2$, 
we obtain, after calculations of integral over $t$ for $n=1$ and $n=2$,
\bea
&& \omega_3({\cal A}, {\cal F})=
\tr \left({\cal A}{\cal F}-\frac13{\cal A}^3\right), \label{3.8}\\
&& \omega_5({\cal A}, {\cal F})=
\tr \left({\cal A}{\cal F}^2-\frac12{\cal A}^3{\cal F}+
\frac1{10}{\cal A}^5\right). \label{3.9}
\eea
Equations (\ref{3.8}) and (\ref{3.9}) have the same forms as the ordinary
Chern-Simons forms for $n=2$ and $n=3$, respectively but it should be
noted that the generalized gauge field ${\cal A}$ and field strength
${\cal F}$ are expressed in matrix forms and 
contain the scalar field as in Eqs.(\ref{2.1}) and (\ref{2.7}).
\par
Let us investigate in more detail the generalized Chern-Simons
form $\omega_{2n+1}$ for $n=1$ and $n=2$.
Inserting ${\cal A}$ and ${\cal F}$ in Eqs.(\ref{2.1}) and (\ref{2.7})
into Eq.(\ref{3.8}), we find
\be
\omega_3=\omega_3^1+\omega_3^2+\omega'_3\,\chi_2\wedge\chi_1,
\label{3.10}
\ee
where
\bea
&&\omega_3^i=\tr\left(A_iF_i-\frac13A_i^3\right),\nonum
&&\omega'_3=\tr\left({\cal D}H_{12}\cdot H_{21}-H_{12}\cdot{\cal D}H_{21}
            \right).
\label{3.11}
\eea
$\omega_3^i(i=1,2)$ is the Chern-Simons form for the ordinary gauge field.
$\omega'_3$ is the new type Chern-Simons form containing the scalar
field $H_{12}$. The transgression formula for this Chern-Simons form 
$\omega'_3$ is written as
\be
d\,\omega'_3=2\tr\left(F_1H_{12}H_{21}-H_{12}F_2H_{21}
                 -{\cal D}H_{12}\cdot{\cal D}H_{21}\right).
\label{3.12}
\ee
When $A_2=0$, Eq.(\ref{3.12}) leads to
\be
d\;\tr\left({\cal D}H_{12}\cdot H_{21}-H_{12}\cdot{\cal D}H_{21}\right)
=2\tr\left(F_1H_{12}H_{21}
                 -{\cal D}H_{12}\cdot{\cal D}H_{21}\right),
\label{3.13}
\ee
where ${\cal D}H_{12}=dH_{12}+A_1H_{12}$ 
and ${\cal D}H_{21}=dH_{21}-H_{21}A_1$. In this case, 
$H_{12}$ belongs to the fundamental representation of the group $G_1$.
Inserting ${\cal A}$ and ${\cal F}$ in Eqs.(\ref{2.1}) and (\ref{2.7})
into Eq.(\ref{3.9}), we find
\be
\omega_5=\omega_5^1+\omega_5^2+\omega'_5\,\chi_2\wedge\chi_1,
\label{3.14}
\ee
where
\bea
&&\omega_5^i=\tr\left(A_1F_i^2-\frac12A_iF_i+\frac1{10}A_i^5\right),\nonum
&&\omega'_5=3\tr\left(F_1{\cal D}H_{12}\cdot H_{21}\right)
            -3\tr\left(F_2{\cal D}H_{21}\cdot H_{12}\right)\nonum
&&\hskip1cm -\frac32d\;\tr\left(F_1H_{12}H_{21}\right)
            +\frac32d\;\tr\left(F_2H_{21}H_{12}\right)
            -\frac12d\;\tr(A_1H_{12}A_2H_{21})
            \nonum
&&\hskip1cm
-\frac12d\;\tr\{A_1({\cal D}H_{12}\cdot
 H_{21}-H_{12}\cdot{\cal D}H_{21})\}  
+\frac12d\;\tr\{A_2({\cal D}H_{21}\cdot H_{12}-H_{21}\cdot{\cal D}H_{12})\}.
\label{3.15}
\eea
$\omega_5^i(i=1,2)$ is the Chern-Simons form for the ordinary gauge field.
$\omega'_5$ is the new type Chern-Simons form containing the scalar
field $H_{12}$. The transgression formula for this Chern-Simons form 
$\omega'_5$ is written as
\bea
&&d\,\omega'_5=3\tr\left(F_1^2H_{12}H_{21}
                 -F_1{\cal D}H_{12}\cdot{\cal D}H_{21}\right)
                 -3\tr\left(F_2^2H_{21}H_{12}
                 -F_2{\cal D}H_{21}\cdot{\cal D}H_{12}\right).
\label{3.16}
\eea
We can find from this equation much more compact equation that
\be
d\,\tr F_1{\cal D}H_{12}\cdot H_{21}=\tr\left(
   F_1^2H_{12}H_{21}-F_1H_{12}F_2H_{21}-F_1{\cal D}H_{12}\cdot{\cal D}H_{21}
  \right),
\label{3.17}
\ee
and the same equation replacing 1 by 2 and 2 by 1 also follows.
When $A_2=0$, Eq.(\ref{3.16}) leads to
\be
d\,\tr\left(F_1{\cal D}H_{12}\cdot H_{21}\right)
=\tr\,\left(F_1^2H_{12}H_{21}
                 -F_1{\cal D}H_{12}\cdot{\cal D}H_{21}\right).
\label{3.18}
\ee
These transgression formulas are easily justified by the direct
calculations according to algebraic rules in the differential
geometry.

\section{Generalized descent equation}
In order to introduce the generalized descent equation \cite{Zumino},
 we incorporate 
the ghost field  \cite{FPS} in the generalized gauge field ${\cal A}$.
\be
{\cal A}^C(x,\theta)={\cal A}(x,\theta)+C(x,\theta)
=\left(\matrix{A_1(x,\theta)+C_1(x,\theta)d\theta & H_{12}(x,\theta)\chi_2\cr
        H_{21}(x,\theta)\chi_1   & A_2(x,\theta)+C_2(x,\theta)d\theta \cr}
                             \right),
\label{4.1}
\ee
where $C_1(x,\theta)$ and $C_2(x,\theta)$ 
are ghost fields belonging to the adjoint
representation of the groups $G_1$ and $G_2$, respectively and 
$\theta$ is an argument of Grassmann number in ghost space. 
$C_i(x,\theta)(i=1,2)$ is anti-Hermitian.
The generalized field strength for ${\cal A}^C$ is given as
\be
{\cal F}^C={\boldsymbol d}{\cal A}^C+{\cal A}^C\wedge{\cal A}^C,
\label{4.2}
\ee
where 
\begin{eqnarray}
&& {\boldsymbol d}=d+d_\theta,\hskip0.5cm d=\partial^\mu dx_\mu,\hskip0.5cm
d_\theta=\partial_\theta d\theta,
\nonumber\\
&&  dx_\mu\wedge d\theta=-d\theta\wedge dx_\mu,
\hskip0.5cm\chi_i\wedge d\theta=-d\theta\wedge\chi_i, \nonumber\\ 
&&d\theta\wedge d\theta\ne0,\hskip0.5cm\partial_\theta^2=0.
\label{4.3}
\end{eqnarray}
Therefore, it is easy to see 
that the exterior derivative ${\boldsymbol d}$ satisfies the nilpotency
${\boldsymbol d}^2=0$.
According to the nilpotency of ${\boldsymbol d}$ and Eq.(\ref{4.3})
, the Bianchi Identity for ${\cal F}^C$
\be
 {\boldsymbol d}{\cal F}^C+[{\cal A}^C,{\cal F}^C]=0
\label{4.4}
\ee
follows. By applying the horizontarity condition \cite{HC} to ${\cal F}^C$,
\be
\left.{\cal F}^C(x,\theta)\right|_{\theta=0}=F(x)
\label{4.5}
\ee
we find the BRST transformations for fields involved.
\bea
&&\delta_\theta A_i=dC_i+A_iC_i-C_iA_i={\cal D}C_i,\nonum
&&\delta_\theta H_{12}=H_{12}C_2-C_1H_{12},\nonum
&&\delta_\theta C_i=-C_i^2,
\label{4.6}
\eea
where the operator $\delta$ stands for the BRST transformation.
\par
According to the nilpotency of ${\boldsymbol d}$ and the Bianchi Identity
for ${\cal F}^C$ in Eq.(\ref{4.4}), we obtain 
the transgression formula for ${\cal A}^C$ and ${\cal F}^C$
 same as in Eq.(\ref{3.6}).
\be
 \tr {{\cal{F}}^C}^{n+1}={\boldsymbol d}\,\omega_{2n+1}
 ({\cal A}^C, {\cal F^C}),
\label{4.7}
\ee
If we consider the horizontarity condition \cite{HC}in Eq.(\ref{4.5}), 
the equation
\be
\left.{\boldsymbol d}\,\omega_{2n+1} 
({\cal A}^C, {\cal F^C})\right|_{\theta=0}=
d\,\omega_{2n+1}({\cal A}, {\cal F})
\label{4.8}
\ee
follows. Here,
\be
\omega_{2n+1}({\cal A}^C, {\cal F}^C)=
(n+1)\int_0^1dt  {\cal A}^C{{\cal F}^C}_t^n,
\label{4.9}
\ee
where
\be
{\cal F}^C_t=t{\boldsymbol d}{\cal A}^C+t^2{\cal A}^C\wedge{\cal A}^C.
\label{4.10}
\ee
By use of Eq.(\ref{4.1}), we expand $\omega_{2n+1}({\cal A}^C, {\cal F}^C)$
in power of the ghost field $C$ as
\be
\omega_{2n+1}({\cal A}^C, {\cal F}^C)=
\omega^0_{2n+1}+\omega^1_{2n}+\omega^2_{2n-1}+\cdots+\omega^{2n+1}_0,
\label{4.11}
\ee
where
the superscript of $\omega$ in the right hand side stands for the power of
the ghost field $C$ and the subscript stands for 
the degree of the form $dx_\mu$.
Here, $\omega_{2n+1}^0$ is the Chern-Simons form.
From Eq.(\ref{4.8}), we find the generalized descent equation
\bea
&&d_\theta\,\omega^0_{2n+1}+d\,\omega^1_{2n}=0,\nonum
&&d_\theta\,\omega^1_{2n}+d\,\omega^2_{2n-1}=0,\nonum
&&d_\theta\,\omega^2_{2n-1}+d\,\omega^3_{2n-2}=0,\\
&&\hskip1cm:\nonum
&&d_\theta\,\omega^{2n}_{1}+d\,\omega^{2n+1}_{0}=0,\nonum
&&d_\theta\,\omega^{2n+1}_{0}=0.\nonumber
\label{4.12}
\eea
$\omega^1_{2n}$ is a solution of the Wess-Zumino consistency condition
\cite{WessZumino}
\be
\Delta=\int \Omega, \hskip1cm d_\theta \Omega=d{\cal B},
\label{4.12a}
\ee
where $\Delta$ is anomaly term, $\Omega$ is a 4-form with ghost number 1
and ${\cal B}$ is a 3-form.
$\omega^1_{2n}$  is written as
\be
\omega^1_{2n}=n(n+1)\int_0^1dt(1-t){\hbox{\rm {STr}}}
(Cd\,({\cal A}{\cal F}_t^{n-1})),
\label{4.13}
\ee
where ${\hbox{\rm {STr}}}$ stands for the symmetrized trace.
We obtain for $n=1,2,3$
\bea
&&\omega^1_2=\tr(Cd\,{\cal A}),\nonum
&&\omega^1_4=\tr\left\{Cd\,\left({\cal A}\,d\,{\cal A}+\frac12{\cal A}^3\right)
\right\},\nonum
&&\omega^1_6=\tr\left\{Cd\,\left({\cal A}\,d\,{\cal A}\cdot d\,{\cal A}
+\frac35{\cal A}\,d\,{\cal A}\cdot{\cal A}^2
+\frac35{\cal A}^3d\,{\cal A}\cdot{\cal A}^2+\frac25{\cal A}^5\right)
\right\},
\label{4.14}
\eea
where the notation of the wedge product is abbreviated.
It should be noted that these equations are same in form as in 
the ordinary case without the scalar field. However, the gauge field
${\cal A}$ written in Eq.(\ref{2.1}) includes the scalar field $H_{12}$.
We extract the term containing the scalar field from Eq.(\ref{4.14}). 
$\omega_2^1$ does not include the scalar field.
\bea
{\omega_4^1}'&=&\tr C_1\,d\,\left\{-H_{12}\,d\,H_{21}+
\frac12\left(A_1H_{12}H_{21}-H_{12}A_2H_{21}+H_{12}H_{21}A_1\right)\right\}
\nonum
&&-\tr C_2\,d\,\left\{-H_{21}\,d\,H_{12}+
\frac12\left(A_2H_{21}H_{12}-H_{21}A_1H_{12}+H_{21}H_{12}A_2\right)\right\}.
\eea
$\omega_6^1$ has the complicated terms containing the scalar field and thus,
we write it in the case of $A_2=C_2=0$.
\bea
{\omega_6^1}'&=&\tr C_1\,d\,\left\{-A_1d\,H_{12}\cdot d\,H_{21}
  {\vphantom{\frac12}}-H_{12}\cdot d\,H_{21}\,d\,A_1\right. \nonum
&&+\frac35\left(A_1\,d\,A_1\cdot H_{12}H_{21}+A_1\,d\,H_{12}\cdot H_{21}A_1
+H_{12}\,d\,H_{21}A_1^2\right)\nonum
&&+\frac35\left(A_1 H_{12}H_{21}\,d\,A_1+H_{12}\cdot H_{21}A_1\,d\,A_1
-A_1^2H_{12}\,d\,H_{21}\right)\nonum
&&+\left.\frac25\left(A_1^3H_{12}H_{21}+A_1^2H_{12}H_{21}A_1+H_{12}H_{21}A_1^3
+A_1H_{12}H_{21}A_1^2\right)\right\}.
\eea
\section{Concluding remarks}
From the standpoint of NCG that the Higgs field is a kind of gauge
field on the discrete space, we incorporated the scalar field $H_{12}$
into the generalized gauge field so as to
generalize the Chern-Simons form and descent equations. 
We obtained the generalized Chern-Simons form and its transgression formula
which include the scalar field $H_{12}$, for example,
as in Eqs.(\ref{3.15}) and (\ref{3.16}).
 The more compact transgression formula follows in Eq.(\ref{3.17}).
We also introduced the generalized descent equations in Eq.(\ref{4.12}).
The physical implications of these formulas will be explored in future
work.

\vskip 0.5 cm
\begin{center}
{\bf Acknowledgement}
\end{center}
\smallskip
The author would like to
express his sincere thanks to
Professor
 H.~Kase and Professor K. Morita 
for useful suggestion and
invaluable discussions. He is grateful to all members 
at Department of Physics, Boston University for their
warm hospitality.
\def\jmp{J.~Math.~Phys.$\,$}
\def\pl{Phys. Lett.$\,$ }
\def\np{Nucl. Phys.$\,$}
\def\ptp{Prog. Theor. Phys.$\,$}
\def\prl{Phys. Rev. Lett.$\,$}
\def\pr{Phys. Rev. D$\,$}
\def\mp{Int. Journ. Mod. Phys.$\,$ }

\end{document}